# EXACT LAW of LIVE NATURE

**Mark Ya.   Azbel'**

**School of Physics and Astronomy, Tel-Aviv University,**

**Ramat Aviv, 69978 Tel Aviv, Israel[+];**

**Max-Planck-Institute für Festkorperforschung – CNRS,**

**F38042 Grenoble Cedex 9, France**

## Abstract

Exact law of mortality dynamics in changing populations and environment is derived. The law is universal for all species, from single cell yeast to humans. It includes no characteristics of animal- environment interactions (metabolism etc) which are a must for life. Such law is unique for live systems with their homeostatic self-adjustment to environment. Its universal dynamics for all animals, with their drastically different biology, evolutionary history, and complexity, is also unique for live systems-cf different thermodynamics of e.g. liquids and glasses. The law which is valid for all live, and only live, systems is a life specific law of nature.

Mortality is an instrument of natural selection and biological diversity. Its law which is preserved in evolution of all species is a conservation law of mortality, selection, evolution, biology. The law implies new kind of mortality and adaptation which dominate in evolutionary unprecedented protected populations and, in contrast to species specific natural selection, proceed via universal stepwise rungs. The law demonstrates that intrinsic mortality and at least certain aspects of aging are disposable evolutionary byproducts, and directed genetic and/or biological changes may yield healthy and vital Methuselah lifespan. This is consistent with experiments. Universality implies that single cell yeast may provide a master key to the cellular mechanism of universal mortality, aging, selection, evolution, and its regulation in all animals. One may look for its manifestations in animal cells also, e.g., in their replicative senescence.

Arguably, universal biology emerged in response to major mass extinctions which posed universal threat to different species, and is related to disposable genes, which were beneficial for longevity in the wild, but became detrimental in evolutionary unprecedented conditions.

Further theoretical and experimental studies of the universal law and its implications are suggested.

**Motivation and approach.** Life evolved via selection of the fittest. Selection posed different challenges for different species, thus yielded biological diversity and complexity of survivors. In contrast, five major mass extinctions [1] were universal "rapidly adjust or die" threat to the very existence of large proportion of species (96% perished in the most drastic extinction about 248 to 238 million years ago). Universal threat could yield certain universality in selection. Indeed, presented physical approach unravels universality which underlies enormous diversity of evolutionary branches. Evolutionary data are sparse and largely qualitative. So, study universality of diverse living species. Selection proceeded via death of the frail. Thus, quantify selection with mortality. To amplify universality, consider different human [2] and protected laboratory populations of med- and fruitflies [3, 4], nematodes (including mutants and biologically amended) [5-8], yeast [5, 9, 10] in changing conditions. Their protection from elements of nature, predators, shortage of resources, diseases, etc nearly eliminates extrinsic mortality, and thus selection, which dominate in the wild. Their predominantly intrinsic mortality is well quantified. It is heterogeneous and non-stationary (e.g., within human lifespan infant mortality increased 30-fold and life expectancy by 50%). Laboratory animal populations (especially genetically homogeneous) are relatively small, and their mortality significantly fluctuates[1]. So, consider life expectancy e and probability $l$ for a

-------------------------------------------------------------------------------------

1. The lifespan of four populations of 623, 662, 248 and 5751 inbred 3X3 male fruitflies in 4-dram shell vials with weekly transfer to fresh medium [3] varied from 18.6 to 34.3 days. In the populations with close life spans (18.6 and 22 days) the probabilities to die on the 38-th day were 18 times different. In the largest population, mortality rate of 15 days olds was 17 times lower than of those 4 days younger and 3 days older. Such giant fluctuations may be related to vial difference and their weekly change. Nematode populations [5-8] include only 50-100 worms.

---------------------------------------------------------------------------------------------------

live newborn to survive to a given age x. These quantities are robust to heterogeneity, non-stationarity and fluctuations. Indeed, suppose the population consists of the groups with the number $N^G(x)$ of survivors to age x. If $C_G = N^G(0)/N(0)$ and $l^G = N^G(x)/N^G(0)$ are correspondingly the ratio of the population and the survivability to x in the group G, then the population survivability $l$ self-averages over population heterogeneity:

$$l = \Sigma N^G(x)/\Sigma N^G(0) = \Sigma C_G\, l^G = <l^G> \tag{1}$$

($<\ldots>$ denote averaging). Since $l = p(0)p(1)\ldots p(x-1)$, where $p(x)$ is the probability to survive from x to $(x+1)$, so $\ln l(x)$ averages $\ln p$ over, and fluctuations with, time x. Similarly, life expectancy $e = <e^G>$ averages over population heterogeneity and entire lifespan. Life expectancy changes 20,000 times from yeast to humans. To account for this change, scale age x and life expectancy e with a single species specific parameter F. Chose F=1 year for humans, F=0.5 day for flies and nematodes, F=0.25 generation for yeast (the choice of F see later). Then Fig. 1 for all animals manifests predominantly universal[2] dependence of survivability $l$ on the scaled life expectancy e/F and age x/F.

---------------------------------------------------------------------------------------------------

2. Actuary Gompertz [11] in 1825 presented the first universal law of mortality for human advanced age. Thereafter the search for such law for all animals went on-see [12, 13] and refs. there. Accurate knowledge of human mortality is important for economics, taxation, insurance, gerontology, etc. So, demographic life tables present millions of mortality data in different countries over their history. To better estimate and forecast mortality, demographers dropped the universal law and developed over 15 mortality approximations [14]. Yet, 180 years after Gompertz, the existence of the universal mortality law remains controversial.

---------------------------------------------------------------------------------------------------

In all cases total survivability $l = l*+ l$ ', where $l*$ is universal, i.e. depends on e/F and x/F only, while non-universal $l$ ', which depends on all multiple factors affecting mortality, is $<<l*$. (From now on, unless specified otherwise, only universal variables are considered, and the superscript star is omitted).

**Universality law: derivation.** Universality for different heterogeneous and homogeneous populations implies that the relation between universal $l$ and e is the same as the relation between their values in any of the groups in the population, i.e. if $l=l(e/F, x/F)$, then $l^G = l(e^G/F, x/F)$. So, by Eq. (1), $l=<l^G>=<l(e^G/F, x/F)>$ and $l= l(e/F, x/F)=l(<e^G/F>, x/F)$, i.e.

$$<l(e^G/F, x/F)>=l(<e^G/F>, x/F) \qquad (2)$$

Such equation implies [15] that $l$ is a universal piecewise linear function of e/F with simultaneous for all ages x/F intersections (denote such dependence as the universal law) and that at any age population heterogeneity, i.e. $e^G/F$ in all groups, is restricted to a single interval ("echelon") of e/F between universal intersections (denote this as a restricted heterogeneity). The universal law agrees with Fig. 1, and restricted heterogeneity implies that dominant fraction of all its populations reduces to a single echelon.

The knowledge of exact analytical dependence on e allows one to establish species specific scales F which provide its minimal relative mean square deviation from experimental[3] data. These scales demonstrate (see Fig. 2) proximity of human (e/F=84)

---------------------------------------------------------------------------------------------------

3. The number of human data (whose statistics is by far the best) included in the approximations was chosen equal to the average number of data per each animal class; human e/F were chosen equidistant. Some scaled ages x/F in Fig. 1 are slightly different for different classes (e.g. 35 fly days are compared to 73 rather than to 35/0.5=70 human years) to amplify universality. At certain ages some intersections in Fig. 1 are weakly pronounced and unobservable.

.

---------------------------------------------------------------------------------------------------

and yeast (e/F=89) survival curves vs x/F despite their ~20,000 times different life expectancies. Proximity is not always as good as in Fig. 1. Empirical study [16] demonstrates very different age dependencies in different echelons (especially in young and old age), thus high sensitivity to contributions of few echelons. Ages and conditions with low mortality may be more universal (thus high e/F in Fig. 2). Poor animal statistics does not allow to account for more than a single echelon. Large size and by far better statistics of human populations allow for it.

Consider period probability d(x)=[ $l$(x) − $l$(x+1)] for a live newborn to die between x and (x+1) years (note that human F=1 and x/F=x). Similar to $l$(x), the value of d(x) self-averages over heterogeneity, but it is more time specific than $l$. The most time specific variable is "infant mortality" d(0)=q(0) which depends on the time from conception to

x=1 only. Similar to Fig. 1, the dependence of d(x) on q(0) for each human curve  is approximately  piecewise linear, also with 5 (as in Fig.1) intersections (see, e.g. Fig. 3) which are nearly simultaneous at all ages, but somewhat different in different countries.  . Since both d(x) and q(0) are  self-averaging variables, previous analysis yields the universal law. Suppose the universal j-th echelon boundaries are

$$q_j < q^G(0) < q_{j+1} \qquad\qquad\qquad (3)$$

Since mortality is never negative, its ultimate minimum is $q_1 = 0$. An arbitrarily heterogeneous population may be distributed at several intervals, and piecewise linear law reduces d(x) to the sum over its intersection values $d_j(x)$:

$$d(x) = \Sigma c_j d_j(x), \ \ \text{where} \ \ \Sigma c_j = 1 \qquad\qquad\qquad (4)$$

**Results.** The number of population specific concentrations $c_j$ of $d_j(x)$ depends on the heterogeneity of the population. If it reduces to a single echelon, thus to two intersections, then d(0)=q(0) and $d_j(0) = q_j$, $d_{j+1}(0) = q_{j+1}$ yield the universal law  :

$$d(x) = c_j d_j(x) + (1-c_j) d_{j+1}(x); \ \ c_j = [\ q_{j+1} - q(0)]/(\ q_{j+1} - q_j) \qquad\qquad (5)$$

The law maps on coexistence of two phases with the "equations of state" $d_j(x)$ and $d_{j+1}(x)$. If a population reduces to two echelons, thus to three intersections, then, by Eq. (4), d(x) reduces to q(0) and one population specific concentration. Simple algebra proves that intersections in all such populations are situated at universal segments of the universal law or their extensions, and allow one to determine the universal law. This is the case in most developed countries (e. g., in 1948–1999 Austria, 1921–1996 Canada, 1921– 2000 Denmark, 1841–1998 England, 1941–2000 Finland, 1899–1997 France, 1956–1999 West Germany, 1906–1998 Italy, 1950–1999 Japan, 1950–1999 Netherlands, 1896–2000 Norway, 1751–2000 Sweden, 1876–2001 Switzerland).  The resulting

universal law in Fig. 4 is verified   with ~3000 human curves [2] (18 countries, two sexes, ages from 1 to 95)-see, e.g., Fig. 3, where straight lines demonstrate the accuracy of approximations with two echelons. (Deviations are more pronounced when infant mortality significantly changes from one calendar year to another due to wars, epidemics, crop failures etc, and/or is relatively large, i.e. when conditions are insufficiently protected). The accuracy may be improved if all echelons in Fig. 4 are accounted for. General universality in Fig. 1 suggests that (properly scaled) law in Fig. 4 is universal for all protected animal populations. Consider its predictions and implications. (Earlier these results were predicted empirically [15, 16, 13] and derived analytically [17]).

**Discussion and conclusions.** Derived law is universal for all species, from yeast to humans.  At a given age x it depends in Fig. 1 on a single population specific parameter- life expectancy e and a single species specific parameter F. The law includes no characteristics of non-stationary and heterogeneous animal- environment interactions (e.g. via metabolism) which are necessary conditions of life. Such law is specific for live systems with their homeostatic self-adjustment to environment. Its dynamics which is universal for all animals, with their drastically different biology, evolutionary history, and complexity, is also unique for live systems-cf different thermodynamics of, e.g., liquids and glasses. The law which is valid for all live, and only live, systems is a life specific law of nature.

Mortality is an instrument of natural selection and biological diversity. The law which is preserved   in evolution of species from humans to yeast is a conservation law of selection, evolution, and biology. It suggests their universal mechanism which dominates in evolutionary unprecedented protected populations (whose mortality is predominantly intrinsic). Then the contribution of all other mechanisms is either relatively small or indirect, via the universal mechanism. Its universality in all animals implies that single cell yeast may be a master key to it and its regulation.

Universal law demonstrates that species specific natural selection is replaced in protected populations by predominantly universal adaptation of intrinsic survivability to genotypes, phenotypes, life history, environment, etc, via properly scaled life expectancy. Universal adaptation is stepwise and proceeds via universal "ladder" of "rungs" with simultaneous for all ages crossovers. Their number equals the number of major extinctions. (Note that each live species in the course of its entire history survived all extinctions). Less universal extinctions may yield mini-rungs, and possibly punctuated evolution [18]). Universal Fig. 1 establishes universal scale of ages for different species, and suggests that life expectancy in all existing species is restricted to around 100 human years, while minor directed genetic and/or biological changes increase it to the Methuselah 250 (healthy and vital [8]) human years.

When infant mortality vanishes, the universal law yields, according to Fig. 4, zero universal mortality till certain age (~80 years for humans), thus correspondingly low total mortality and high life expectancy. Mortality on the scale of stochastic fluctuations, i.e. consistent with zero universal mortality, was indeed observed in humans, flies, nematodes, yeast. In 2001 Switzerland  only 1 (out of 60,000) girl died at 5, 9, and 10 years; 5 girls died in each age group from 4 to 7 and from 9 to 13 years; 10 or less from 2 till 17 years; no more than 16 from 2 till 26. Statistics is similar in all 1999-2002 Western developed countries [2]. Similarly, mortality of dietary restricted flies at 8 days was ~0.0004 [19].  Yeast mortality [9] was zero during half of its mean life span (Jazwinski et al, 1998 presented the first model which stated that a sufficient augmentation of aging process resulted in a lack of aging). The probability to survive from 80 to 100 years increased in Western Europe 20-fold in the last 50 years [20]. Mean life expectancy increased almost three times in the last 250 years with improving (medical included) human conditions [2]; 2.4-fold with genotype change in Drosophila [3]. None of nematodes with changes in small number of their genes and tissues [8] died till 27 days, i.e. during 54 human years on Fig. 1 scale; from 58 till 90, from 126 till 162 "human years". 25% of amended nematodes survived till 296 and thereafter did not die till 318 "human years". Zero mortality till certain age implies zero universal mortality at any age (unless it has a singularity at certain age [21]), thus very low total mortality, and the Methuselah life expectancy.  Indeed, mean life span of mutant nematodes increased to 90 days [6,7] and to the Methuselah 124 days [8] (248 years on "human" scale), with no apparent loss in health and vitality.

An important implication of the universal law is its plasticity. Universal mortality at any age is related to infant mortality[4] (see Fig. 4). Thus, it  rapidly adjusts to, and is

-------------------------------------------------------------------------------------------------

[4] Thus, eliminating all deaths before age 50 would not yield just about a 4-year rise in current life expectancy at birth, as it would if mortality at higher ages were little correlated with lower age mortality. Demo- and biodemographers consider the most specific mortality variable-the probability to die between ages x and (x+1). It equals $q(x)= [l(x) - l(x+1)]/ l(x )=d(x)/[1-d(0)-d(1)-…-d(x-1)]$, thus  its universality is not as explicit as that of d(x) and was noticed.

-----------------------------------------------------------------------------------------------

determined by, current living conditions if they do not significantly change in 2 years, from conception till 1 year, for humans; few percent of the life span for any species. So, universal mortality is independent of the previous life history ("short mortality memory" of it) and, together with infant mortality, it may be rapidly reduced and reversed to its value at much younger age[4]. Indeed, following unification of East and West Germany, within few years mortality in the East declined toward its levels in the West, especially among elderly with ~45 years of their different life histories[4]. Mortality of the female cohort, born in 1900 in neutral Norway, at 59 years restored its value at 17 years, i.e. 42 years younger [2]. Note that such mortality decrease, similar to the one in East Germany after its unification, is not dominated by death of the frail. The latter alters composition of the cohort, and the resulting change in mortality depends on life history rather than on current conditions only. Thus, it contributes to the deviations from the universal laws (which are relatively small) rather than to the universal mortality. Mortality plasticity is also very explicit in experiments where dietary restriction in rats [22] and flies [19] is switched on. However, when dietary restriction is switched off and changes to full feeding, their longevity remains higher than in the control group of animals fully fed throughout life. Also, when fly temperature was lowered from 27 to 18 degrees or vice versa, the change in mortality, driven by life at previous temperature, persisted in these flies compared to the control ones. Such long memory of life history may be related to rapid changes in temperature or feeding, since universal law is valid when infant mortality little changes within a day for flies, a month for rats, a year for humans. This calls for comprehensive tests of mortality adaptation to such conditions. Similar tests may verify a possibility to reverse and reset mortality of a homogeneous cohort to a much younger age.

Restricted heterogeneity implies that at the intersections population homogenizes. This agrees with experimental data [15].

**Outstanding problems.** Vanishing and highly plastic universal mortality calls for evolutionary and biological explanation. In the wild competition for sparse resources is fierce, and only relatively few genetically fittest animals survive to their evolutionary "goal"- reproduction. Even human life expectancy at birth was around 40-45 years just over a century ago, and 17.2 years for males in 1773 (crop failure year) Sweden [2]. There are no evolutionary benefits from genetically programmed death and/or aging of tiny number of survivors to old age. Since  prior to and during reproductive age (when survival is evolutionary beneficial) mortality, and even aging (thus irreparable damage also), may be negligible in protected populations (see above); since there are no evolutionary benefits in switching off repair mechanisms later, so intrinsic mortality and aging are presumably disposable evolutionary byproducts. Such byproducts may be related to genes, which are beneficial for non-universal longevity in the wild, but are detrimental in evolutionary unprecedented protected conditions where longevity is predominantly universal (new kind of Williams antagonistic pleiotropy). "Byproduct" genes are relatively easy to alter or switch off.  This is consistent with healthy and vital Methuselah age in nematodes. Universality suggests that its mechanism  may be reduced to genetically regulated universal processes in cells, and related to a certain universal genome (cf "longevity genes" [23-26]). Single cell yeast may provide a master key to the cellular mechanism of Methuselah age, adaptation, and their regulation, in all animals (see cartoon in Fig. 5). One may look for its manifestations in animal cells, e.g., in their replicative senescence (see review [27] and refs. therein), apoptosis, possibly even in certain aspects of cancer ([28] and refs. therein) and cancer gene therapy (e.g., inhibition of ontogenes and activation of tumor suppressor genes).

In protected populations non-universal mortality is relatively small, thus all its mechanisms are less important than or correlated with universal mechanism.

Conservation law of universal evolution allows for its quantitative study with current survivors, as well as for accurate definition of species, families etc according to their scales in Figs. 1 and 4. Remarkably simple scales in Fig. 1 suggest the existence of their "quantization law".

Universal piecewise linear dependence on e/F is related to its invariance to restricted population heterogeneity. Invariance which yields analytical formula of the universal dependence on age, remains to be established, as well as accurate species specific scaling of mortality dependence on age and infant mortality.

Universal law presents universal demo- and biodemographic approximation, which may be important for economics, taxation, insurance, gerontology, etc.

Interconnection between universal evolution, selection, mortality, aging and its vitality, and presumably mass extinctions, suggests certain universality in biology at large and calls for multidisciplinary (evolutionary, biological, demographic, physical and mathematical) study. Universal law, its implications, and predictions may be comprehensively verified and refined theoretically (with available mortality data) and experimentally. Other outstanding problems include "quantization law" of evolutionary scales in Fig. 1, 4; genes and cellular mechanism of universal mortality; physical and biological nature of intersections and echelons.

Acknowledgement.  I am very grateful to I. Kolodnaya for assistance and cartoon. Financial support from A. von Humboldt award and R. & J. Meyerhoff chair is highly appreciated.

APPENDIX

Demographic life tables present  mortality data in different countries over their history . For males and females, who died in a given country in a given calendar year, the data list, in particular, "period" probabilities q(x) (for survivors to x) and d(x) (for live newborns) to die between the ages x and (x+1) [note that d(0)=q(0)]; the probability $l$(x) to survive to x for live newborns; the life expectancy e(x) at the age x. The tables also present the data and procedures which allow one to calculate the values of q(x), d(x), $l$(x), e(x) for human cohorts, which were born in a given calendar year.  Populations, their conditions and heterogeneity are different, yet demographic approximations reduce period mortality of any given population to few parameters. Assumption that under certain conditions a dominant fraction of period mortality in all heterogeneous populations is universal is sufficient to derive the universal Fig. 1, as well as Eq.(5) and its conditions (3). According to Fig.4, until $\sim$ 65 years, d(x) decreases when q(0) increases.  Beyond $\sim$ 85 years, d(x) increases together with q(0).  In between, d(x) exhibits a well pronounced maximum (smeared by generic fluctuations). Consider the origin of such dependence on age.  The value of d(x) is proportional to the probabilities for a newborn to survive to x and then to die before the age (x+1).  When living conditions improve, the former probability increases, while the latter one decreases.  In young age the probability to survive to x is close to 1, so d(x) is dominated by the mortality rate, and thus monotonically decreases together with q(0).  For sufficiently old age, low probability to reach x dominates.  It increases with improving living conditions, i.e. with decreasing q(0), thus d(x) increases with decreasing q(0). At an intermediate age, when improving living conditions sufficiently increase survival probability, d(x) increase is replaced with its decrease.  Then d(x) has a maximum at a certain value of q(0).  Thus, minor genetic and/or biological changes should yield the d(x) maximum at 95 years and beyond. To quantify the accuracy of the results, consider the number D(x) of deaths at a given age x in each calendar year. According to statistics, the corresponding stochastic (i.e. minimal) error is $\sim 2/[D(x)]^{1/2}$.  At 10 years of age it increases from ~20% in 1976 to ~200% in 2001 Switzerland and leads to large fluctuations in q(10).  At 40 years it is ~20%; at 80 years it is ~6% in Switzerland and ~2% in Japan.

Universal Fig. 4 and accuracy of the universal d(x) vs d(0) with two echelons may be refined with larger number of echelons in populations.  The total number of equations (4) is 2XT, where 2 is the number of sexes, (X-1) is the maximal considered age, $T = \Sigma \; T_g$, where $T_g$ is the number of calendar years in the period life tables of the country g. The total number of Eq. (4) variables with 5 intersections is 10T+5X. Since T~2000, X~100, the number of variables is~20 times less than the number of equations. So, consider non-universal mortality with the concentrations in Eq. (4)  which change with age (e.g., every five years) to provide the same number of equations and variables. The latter change, calculated according to life tables and Eq. (4), determines relative non-universal mortality.

FIGURE CAPTIONS

Fig. 1. Universal dependence of survivability (vertical axis) on scaled life expectancy e/F (horizontal axis) for given scaled ages x/F of 125 Swiss (1876-2001 years, crosses) and 50 Japanese (1950-1999, dashes) female [2]; 17 fly [3,4] (black) and 15 nematode [5-8] (white); 14 yeast [5, 9, 10] (circles) populations. Their scaling parameters F and ages x are correspondingly F=1 year; 0.5 days; 0.25 generations and x=30, 85, 73 (upper, lower and middle curves) years; 15, 45 (squares), 35 (triangles) days; 7, 21 (white), 16 (black) generations. Each sign presents a population.
Some signs overlap and are indistinguishable for humans and flies, nematodes and yeast. Few accidental deviations are omitted. Solid lines demonstrate the universal law. The difference between presented and all other human data (e.g. those for, e.g., 252 Swedish female and 159 English male populations) is on the scale of difference between nematodes and yeast.

Fig. 2. Survivability vs scaled age for females who died in 1999 Japan [2] (black triangles) and for yeast [5] (white triangles). Their scaled life expectancies are correspondingly 84 and 89.

Fig. 3. (Upper plot). Period probabilities for live newborn Japanese (black) and Swedish (white) females to die (year of death from 1950 to 1999 and 1751 to 2002) between 60 and 61 (squares), 80 and 81 (triangles), 95 and 96 (diamonds) years of age vs. infant mortality q(0). Japanese relative mean squared deviations from the universal law with two echelons  (straight lines) are correspondingly 2.4%, 2.3% and 10%. Significant Swedish deviations are related to 1918 flu pandemic in Europe.

(Lower plot). Same for French (diamonds) and  Japanese (triangles) females (year of death from 1898 and 1950 to 2001 and 1999) between 80 and 81 years of age. Empty diamonds correspond to 1918 flu pandemic and World Wars. They are disregarded in the universal law with two echelons  (straight lines), which yield relative mean square deviations from black signs on the scale of generic 5%. When Japanese $q(0)= 0$, its extrapolation yields $d(80)=0$.

Fig. 4. Universal law (thick lines) of human mortalities $d(60)$, $d(80)$ and $d(95)$ vs $q(0)$-middle, lower and upper  curves. At $q(0)<0.003$ they are extrapolated.   Thin lines extend universal linear segments. Country specific intersections (similar to those in Fig. 3) are exemplified by diamonds and squares for England (two successive intersections), France, Italy and Japan, Finland, Netherlands, Norway, Denmark, France, England. All intersections are close to universal straight lines.

Fig. 5. The ladder of rungs in the human "bridge of death". Better social and medical protection at its successive rungs implies higher "protective walls" against, thus delay in, death and aging, but does not shift the precipice  at the bridge end. Biological amendments increase the maximal life span and shift the "bridge of death" end.

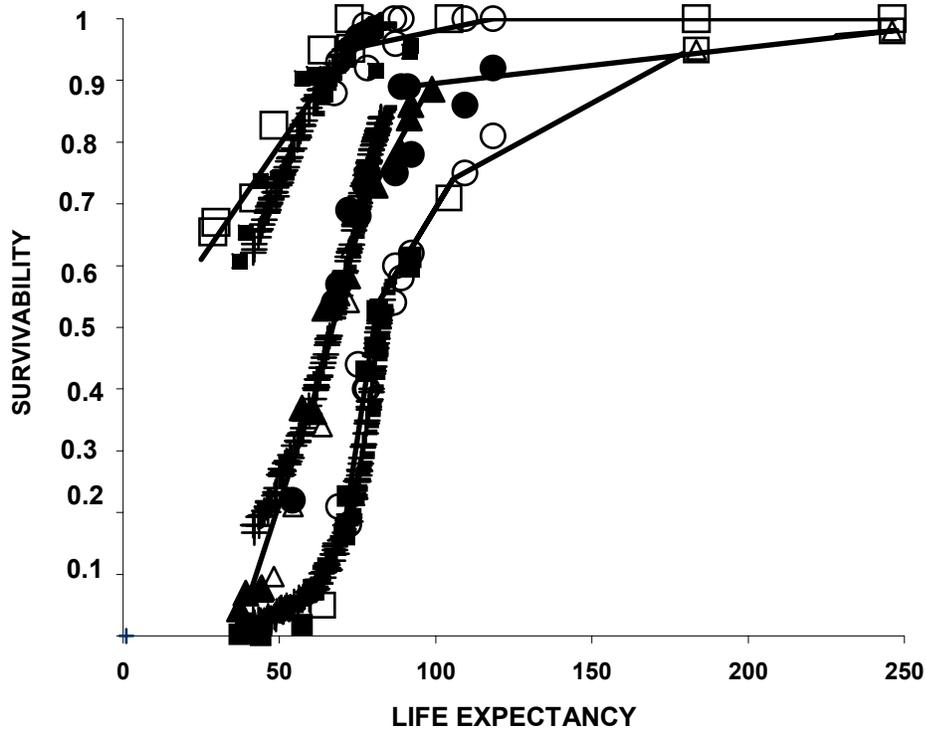

**Fig.1**

SURVIVABILITY

LIFE EXPECTANCY

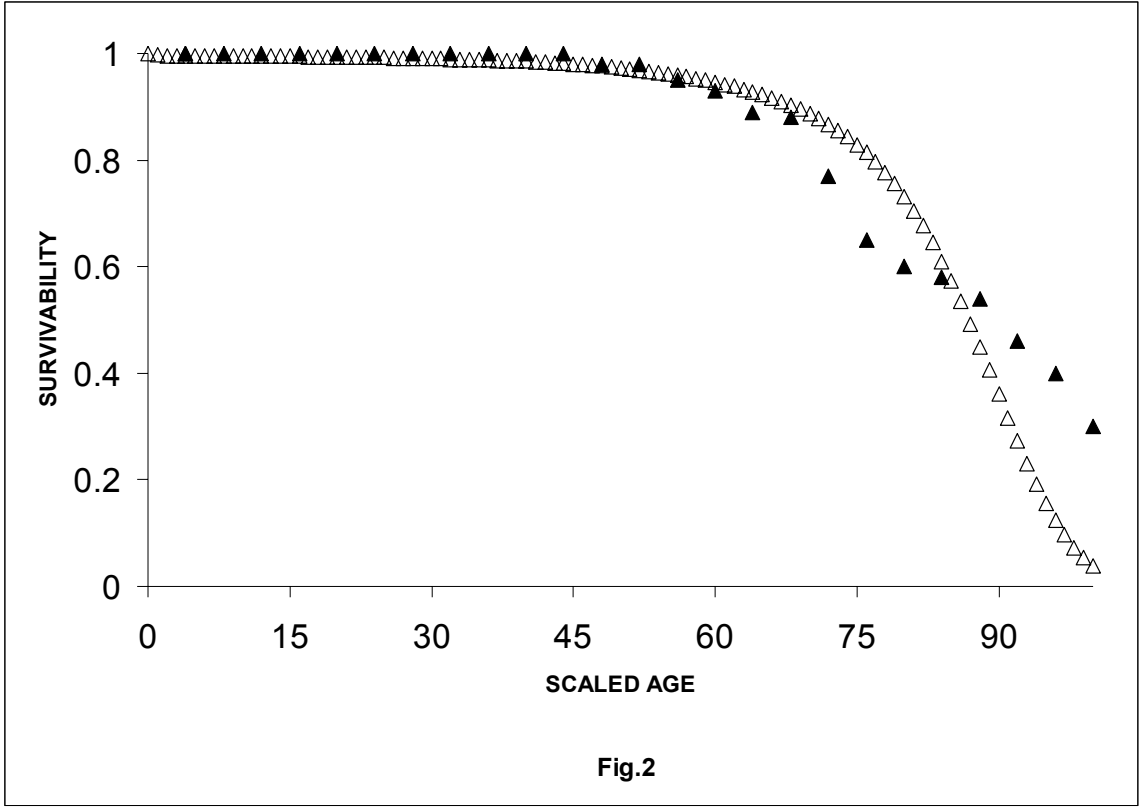

**Fig.2**

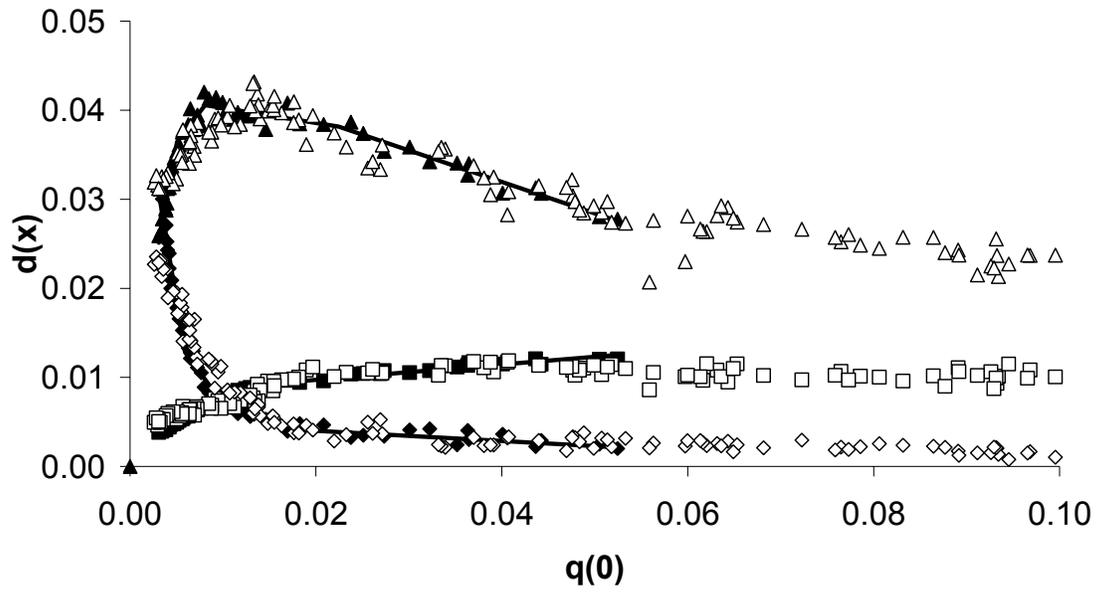

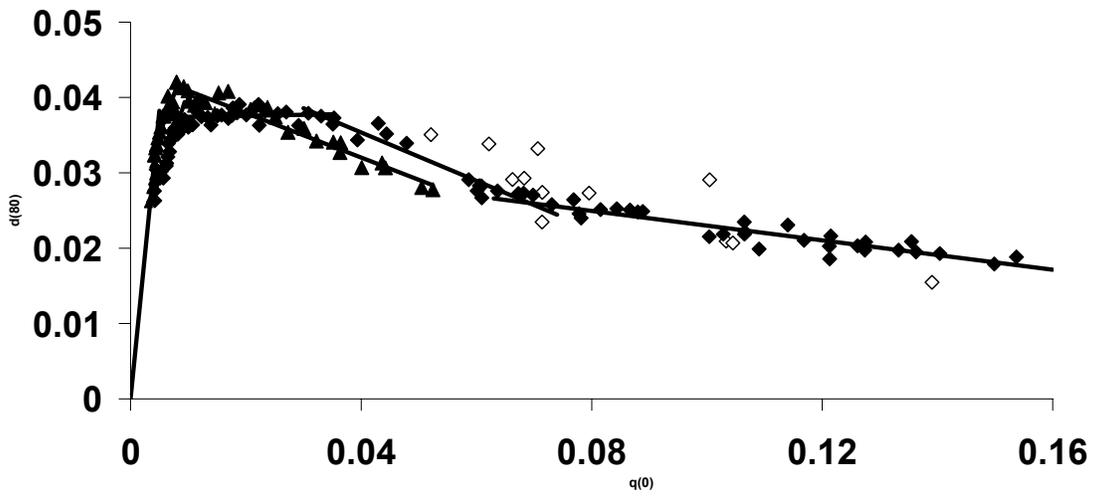

**Fig. 3**

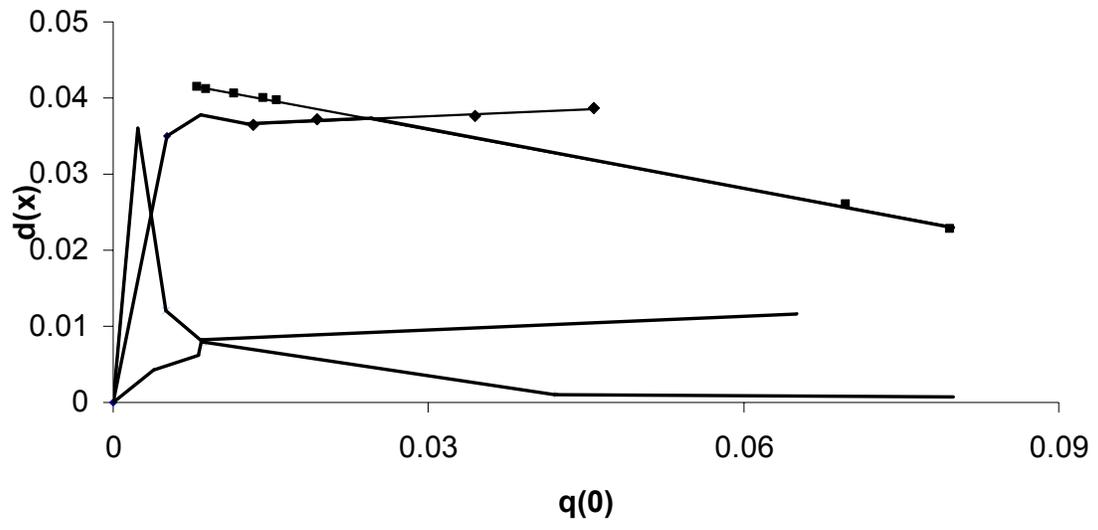

Fig. 4

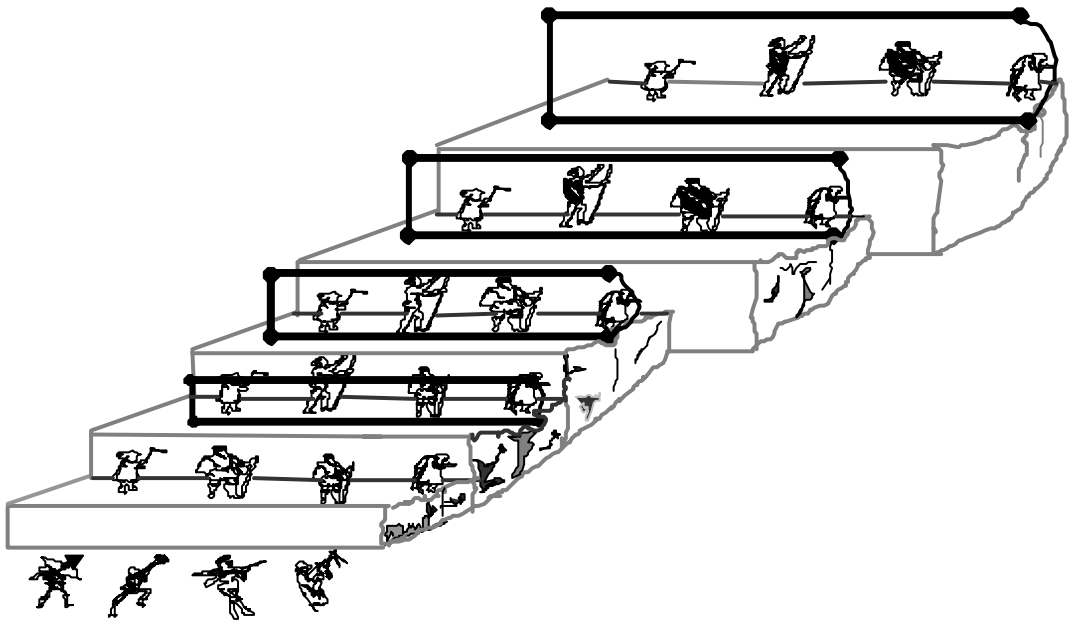

Fig. 5